\documentclass[journal=apchd5,manuscript=article]{achemso}

\usepackage{chemformula} 
\usepackage[T1]{fontenc} 

\usepackage{amssymb}
\usepackage{amsmath}

\author{Nuttawut Kongsuwan}
\affiliation{The Blackett Laboratory, Prince Consort Road, Imperial College London, London~SW7~2AZ, United Kingdom}

\author{Angela Demetriadou}
\affiliation{School of Physics and Astronomy, University of Birmingham, Edgbaston, Birmingham, B15 2TT, United Kingdom}

\author{Matthew Horton}
\author{Rohit Chikkaraddy}
\author{Jeremy J. Baumberg}
\affiliation{Nanophotonics Centre, Cavendish Laboratory, University of Cambridge, Cambridge CB3 0HE, United Kingdom}
\email{jjb12@cam.ac.uk}

\author{Ortwin Hess}
\affiliation{The Blackett Laboratory, Prince Consort Road, Imperial College London, London~SW7~2AZ, United Kingdom}%
\email{o.hess@imperial.ac.uk}

\title[]{Plasmonic nanocavity modes: From near-field to far-field radiation}

\abbreviations{NP,NPoM,QNM}

\begin{document}




\begin{abstract}
In the past decade, advances in nanotechnology have led to the development of plasmonic nanocavities which facilitate light-matter strong coupling in ambient conditions. The most robust example is the nanoparticle-on-mirror (NPoM) structure whose geometry is controlled with subnanometer precision. The excited plasmons in such nanocavities are extremely sensitive to the exact morphology of the nanocavity, giving rise to unexpected optical behaviors. So far, most theoretical and experimental studies on such nanocavities have been based solely on their scattering and absorption properties. However, these methods do not provide a complete optical description of a NPoM. Here, the NPoM is treated as an open non-conservative system supporting a set of photonic quasinormal modes (QNMs). By investigating the morphology-dependent optical properties of nanocavities, we propose a simple yet comprehensive nomenclature based on spherical harmonics and report spectrally overlapping bright and dark nanogap eigenmodes. The near-field and far-field optical properties of NPoMs are explored and reveal intricate multi-modal interactions.
\end{abstract}

\section*{Introduction}
Metallic nanostructures have the ability to confine light below the diffraction limit via the collective excitation of conduction electrons, called localized surface plasmons. Through recent advances in nanofabrication techniques, gaps of just 1-2 nm between nanostructures have been achieved \cite{anger2006enhancement, savage2012revealing}. At such extreme nanogaps, the plasmonic modes of two nanostructures hybridize to allow an unprecedented light confinement \cite{prodan2003hybridization,zhu2016quantum}, making coupled nanostructures an ideal platform for field-enhanced spectroscopy \cite{kinkhabwala2009large,chen2012experimental}, photocatalysis \cite{zhang2013plasmonic} and nano-optoelectronics \cite{choo2012nanofocusing}. One such nanostructure is the nanoparticle-on-mirror (NPoM) geometry where a nanoparticle is separated from an underlying metal film by a molecular mono-layer \cite{ciraci2012probing, mertens2013controlling}. This geometry (which resembles the prototypical dimer but is more reliable and robust to fabricate) has attracted considerable interest since it enables light-matter strong-coupling of a single molecule at room temperature \cite{chikkaraddy2016single}, and it has many potential applications, including biosensing \cite{li2013single} and quantum information \cite{hensen2017strong,ojambati2019quantum}.

A wide range of theoretical and experimental studies have been conducted to investigate the optical properties of NPoM nanocavities \cite{li2013single, mubeen2012plasmonic, baumberg2019extreme}. Several studies examine resonances of NPoMs \cite{sigle2015monitoring, tserkezis2015hybridization, huh2018comparative, demetriadou2017spatiotemporal, devaraj2019modifying}  and their influence on optical emission of single molecules in the nanogaps \cite{kongsuwan2018suppressed, kongsuwan2018fluorescence, chikkaraddy2018mapping}. However, most studies on the nanocavities have so far described their optical response via a scattering method and infer their resonances from resulting far-field spectral peaks. Although significant information can be obtained from the far-field spectra, they do not reveal complete optical descriptions of the nanocavities and commonly hide information about their dark modes. For example, an incident field from the far-field does not couple to all available photonic modes of the system. Resonances which are spectrally close also interfere with each other and often merge into single broadened peaks in far-field spectra. As a result, analyses of the far-field scattered spectra can yield inconsistency between near-field and far-field plasmonic responses \cite{lombardi2016anomalous}. 

The precise morphological details of the NPoM nanogaps also dramatically modify their optical responses \cite{sigle2015monitoring, benz2016sers}. Once the nanoparticles are placed on substrates, they lie on their facets which can have varying widths. Previous studies of the gap morphology of NPoMs described their gap plasmonic resonances with two sets of modes: longitudinal antenna modes, excited for all facet widths, and transverse waveguide modes, excited at large facet widths \cite{tserkezis2015hybridization, huh2018comparative}. Although this description sheds light upon the asymptotic behavior of the NPoM plasmons, it provides an incomplete picture of NPoM resonances at intermediate facet widths.

In this article, we identify and map the inherent resonant states of the NPoM geometry by employing the quasinormal mode (QNM) method \cite{lalanne2018light}. We propose a simple yet comprehensive nomenclature based on spherical harmonics to describe the resonances of NPoMs with varying morphology. A collection of spectrally overlapping bright and dark nanogap QNMs are reported, including some photonic modes that have not been reported elsewhere. These results imply that a quantum emitter placed inside a NPoM nanogap couples to a collection of QNMs and experiences a complex multi-modal interaction. By calculating the far-field emission of each QNM using a near-to-far-field transformation (NFFT) method, we predict the total emission profile of a dipole emitter placed inside a NPoM nanogap from the QNM collective responses. The resulting emission profiles display rich and intricate behaviors, governed by the NPoM morphology.

\section*{Eigenmodes of plasmonic nanocavities}
\label{sec:qnm}

A general open system is non-conservative as its energy leaks to its surrounding environment, and therefore, its time-evolution is non-Hermitian. Consequently, its resonance can no longer be described by a normal mode but instead is characterized by a QNM with a complex eigenfrequency \cite{ching1998quasinormal}. The QNM analysis is a standard methodology to study open and dissipative systems, of which the approximate descriptions are often provided by a few QNMs. This approach has spanned a wide range of applications, including gravitational waves from black holes \cite{kokkotas1999quasi} and electromagnetic waves from nanoresonators \cite{lalanne2018light}.

In nanophotonics, significant progress has been made in the past decades towards solving QNMs for general dispersive materials. Efficient QNM solvers have been  developed using a variety of techniques, including the time-domain approach \cite{ge2014design,kristensen2014modes}, the pole-search approach \cite{bai2013efficient,powell2014resonant,zheng2014implementation} and auxiliary-field eigenvalue approach \cite{yan2018rigorous}. For resonators with arbitrary shapes and materials, analytic solutions are not generally available, and several numerical methods have been developed which surround the resonators by perfectly matched layers (PMLs) to approximately simulate infinite domains \cite{sauvan2013theory,yan2018rigorous}.

Here, we represent the resonances as QNMs with complex frequencies $\widetilde{\omega}_{i} = \omega_{i} - i \kappa_{i}$ where the real part $\omega_{i}$ is the spectral position and the imaginary part $\kappa_{i}$ is the linewidth, i.e., dissipation rate. For a general optical system with non-magnetic materials, its QNMs can be found by solving the time-harmonic and source-free Maxwell's equations \cite{lalanne2018light}
\begin{equation}
    \begin{pmatrix} 0 & -i\mu^{-1}_0\nabla\times \\ i\varepsilon(\mathbf{r};\widetilde{\omega}_{i})^{-1}\nabla\times & 0 \end{pmatrix}
    \begin{pmatrix} \widetilde{\mathbf{H}}_{i}(\mathbf{r}) \\ \widetilde{\mathbf{E}}_{i}(\mathbf{r}) \end{pmatrix}
    =
    \widetilde{\omega}_{i}
    \begin{pmatrix} \widetilde{\mathbf{H}}_{i}(\mathbf{r}) \\ \widetilde{\mathbf{E}}_{i}(\mathbf{r}) \end{pmatrix},
\label{eq:qnmMaxwell}
\end{equation}
where $\varepsilon(\mathbf{r};\widetilde{\omega}_{i})$ is the permittivity and $\widetilde{\mathbf{H}}_{i}$ and $\widetilde{\mathbf{E}}_{i}$ are the magnetic and electric fields of a QNM which satisfy the Sommerfeld radiation condition for outgoing waves. 

For dispersive materials like metals, Eq. (\ref{eq:qnmMaxwell}) is, in general, a nonlinear eigenvalue equation. However, Eq. (\ref{eq:qnmMaxwell}) can be converted into a linear equation if the material permittivities are described by an $N$-pole Lorentz-Drude permittivity
\begin{equation}
	\varepsilon(\omega) = \varepsilon_\infty\left(
	    1 
	    - \sum^N_{k=1} \frac{\omega^2_{p,k}}{\omega^2 - \omega^2_{0,k} + i\gamma_k\omega}
	\right)
\label{eq:lorentzDrude}
\end{equation}
where $\varepsilon_\infty$ is the asymptotic permittivity at infinite frequency while $\omega_{p,k}$, $\omega_{0,k}$ and $\gamma_{k}$ are the plasma frequency, resonant frequency and decay rate, respectively, of the $k^\mathrm{th}$ Lorentz-Drude pole. For each Lorentz-Drude pole in the summation, two auxiliary fields can be introduced \cite{yan2018rigorous}
\begin{equation}
    \widetilde{\mathbf{P}}_{i,k}(\mathbf{r}) = -\frac{\varepsilon_\infty\omega^2_{p,k}}{\widetilde{\omega}_{i}^2 - \omega^2_{0,k} + i\gamma_k\widetilde{\omega}_{i}}\widetilde{\mathbf{E}}_{i}(\mathbf{r}),
\end{equation}
\begin{equation}
    \widetilde{\mathbf{J}}_{i,k}(\mathbf{r}) = -i\widetilde{\omega}_{i}\widetilde{\mathbf{P}}_{i,k}(\mathbf{r})
\end{equation}
where $\widetilde{\mathbf{P}}_{i,k}$ and  $\widetilde{\mathbf{J}}_{i,k}$ are auxiliary polarization and current, respectively, of the i$^\mathrm{th}$ QNM and the $k^\mathrm{th}$ Lorentz-Drude pole. In this article, we limit our discussion to gold with permittivity \cite{johnson1972optical} sufficiently described by a two-pole Lorentz-Drude model with $\varepsilon_\infty = 6 \varepsilon_0$, $\omega_{p,1} = 5.37\times 10^{15}$ rad/s, $\omega_{0,1} = 0$ rad/s, $\gamma_{1} = 6.216\times 10^{13}$ rad/s, $\omega_{p,2}= 2.2636\times10^{15}$ rad/s, $\omega_{0,2} = 4.572\times 10^{15}$ rad/s and $\gamma_2 = 1.332\times 10^{15}$ rad/s. By inserting the auxiliary fields into Eq. (\ref{eq:qnmMaxwell}), we obtain a linear eigenvalue equation
\begin{equation}
    \widehat{\mathcal{H}}\widetilde{\psi}_{i}
    =
    \widetilde{\omega}_{i} \widetilde{\psi}_{i}
\label{eq:qnmMaxwellGold}
\end{equation}
\begin{equation}
    \widetilde{\psi}_{i} = 
    \begin{pmatrix} 
    \widetilde{\mathbf{H}}_{i}(\mathbf{r})     &
    \widetilde{\mathbf{E}}_{i}(\mathbf{r})     & 
    \widetilde{\mathbf{P}}_{i,1}(\mathbf{r}) & \widetilde{\mathbf{J}}_{i,1}(\mathbf{r}) & \widetilde{\mathbf{P}}_{i,2}(\mathbf{r}) & \widetilde{\mathbf{J}}_{i,2}(\mathbf{r}) 
    \end{pmatrix}^T
\end{equation}
\begin{equation}
    \widehat{\mathcal{H}} = 
    \begin{pmatrix} 
    0 & -i\mu^{-1}_0\nabla\times  & 0 & 0 & 0 & 0 \\ i\varepsilon^{-1}_\infty\nabla\times & 0 & 0 & -i\varepsilon^{-1}_\infty & 0 & -i\varepsilon^{-1}_\infty \\
    0 & 0 & 0 & i & 0 & 0 \\
    0 & i\omega^2_{p,1}\varepsilon_\infty & -i\omega^2_{0,2} & -i\gamma_1 & 0 & 0 \\
    0 & 0 & 0 & 0 & 0 & i \\
    0 & i\omega^2_{p,2}\varepsilon_\infty & 0 & 0 & -i\omega^2_{0,2} & -i\gamma_2,
    \end{pmatrix}.
\end{equation}

Unlike real-frequency normal modes, the QNMs have unique properties such that their fields diverge in space at large distance. Hence, the QNMs do not have finite energy and cannot be normalized based on energy consideration \cite{lalanne2018light}. The orthogonality relation and normalization of the QNMs are given by their unconjugated products of the form \cite{sauvan2013theory,lalanne2018light}
\begin{equation}
    \iiint_\Omega\left[
        \widetilde{\mathbf{E}}_{j}\cdot\frac{
            \widetilde{\omega}_{j}\varepsilon(\mathbf{r};\widetilde{\omega}_{j}) -
            \widetilde{\omega}_{i}\varepsilon(\mathbf{r};\widetilde{\omega}_{i})
        }{\widetilde{\omega}_{j}-\widetilde{\omega}_{i}}
        \widetilde{\mathbf{E}}_{i} - 
        \mu_0\widetilde{\mathbf{H}}_{j}\cdot\widetilde{\mathbf{H}}_{i}
    \right] d^3 \mathbf{r} = 0
\label{eq:qnmOrthogonality}
\end{equation}
\begin{equation}
    \iiint_\Omega\left[
        \widetilde{\mathbf{E}}_{i}\cdot\left(\frac{
                \partial[\omega\varepsilon(\mathbf{r};\omega)]
            }{
                \partial\omega
            }\right)_{i}
            \widetilde{\mathbf{E}}_{i}
        - \mu_0\widetilde{\mathbf{H}}_{i}\cdot\widetilde{\mathbf{H}}_{i}
    \right] d^3 \mathbf{r} = 1
\label{eq:qnmNorm}
\end{equation}
where the derivative is evaluated at ${\omega=\widetilde{\omega}_{i}}$ and the integration domain $\Omega$ must include both the system and the PMLs.

\begin{figure}[!ht]
\centering
\includegraphics[width=0.8\columnwidth]{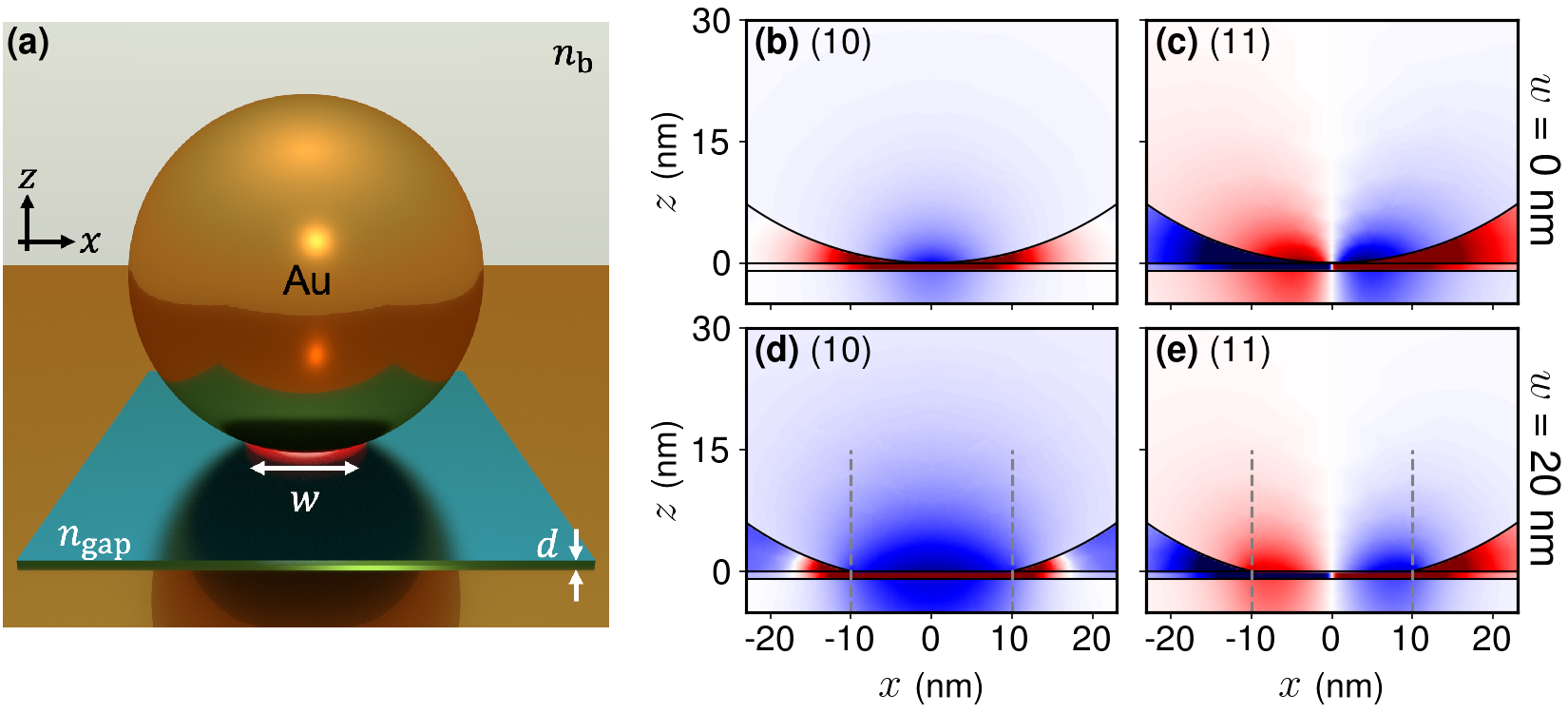}
\caption{(a) Schematic for a gold nanoparticle-on-mirror (NPoM) structure with diameter $2R = 80$ nm, nanogap $d=1$ nm, nanogap index $n_\mathrm{gap}=1.45$, background index $n_\mathrm{b}=1$ and facet width $w$. (b, c, d, e) normalized quasinormal-mode (QNM) electric fields $E_{z,\ell\mathit{m}}$ in the vertical xz-plane of NPoMs with (b, c) $w =$ 0 and (d, e) $w =$ 20 nm for (b,d) the (10) mode and (c, e) the (11) mode. The gray dashed lines indicate the facet edges.}
\label{fig:qnm_E_xz}
\end{figure}

To calculate the QNMs of a gold NPoM structure, we employ the above methodology by using QNMEig, an open-source program based on COMSOL Multiphysics, which implements an efficient finite-element solver by accounting for material dispersion with auxiliary fields \cite{yan2018rigorous}. The following parameters are chosen based on recent experiments \cite{chikkaraddy2016single, sigle2015monitoring}, with nanoparticle diameter of $2R=80$ nm, gap thickness $d=1$ nm, refractive index $n_\mathrm{gap}=1.45$ and background refractive index $n_\mathrm{b}=1.0$, as shown in Fig. \ref{fig:qnm_E_xz}(a).

Here, we introduce a comprehensive nomenclature based on spherical harmonics $Y^\mathit{m}_{\ell}$. Each QNM is labelled with $i=(\ell\mathit{m})$ where $\ell =$ 0, 1, 2, ... and  $-\ell\leq \mathit{m} \leq \ell$. For an isolated spherical nanoparticle, spherical harmonics form a complete set of orthogonal basis functions for its plasmonic modes. For a NPoM with a spherical nanoparticle, these plasmonic modes are coupled to their image charges in the mirror and become densely distributed near the gap, forming gap plasmons. We report here that the gap plasmons in the nanocavity preserve the nomenclature of plasmons of an isolated nanoparticle despite the fact that the actual plasmons are highly deformed by the mirror.

Figure \ref{fig:qnm_E_xz}(b, c) shows the real z-component of QNM fields $E_{z,\ell\mathit{m}}=\mathrm{Re}[\widetilde{\mathbf{E}}_{\ell\mathit{m}}\cdot \hat{\mathbf{e}}_z]$ for the two lowest-eigenfrequency QNMs, denoted as (10) and (11), of a spherical NPoM ($w =$ 0 nm). At the nanogap, these QNMs exhibit large field confinement far below the diffraction limit, which is the main characteristic of gap plasmons. The (10) mode, also commonly referred to as the bonding dipole plasmon, is symmetric (even) across the $x=0$ plane whereas the (11) mode is antisymmetric (odd), which also reveal their bright and dark nature respectively. Note that for electromagnetic problems, the (00) mode is undetermined and does not contribute to the fields \cite{bohren1998absorption, low1997classical}. 

\begin{figure}[!ht]
\centering
\includegraphics[width=0.9\columnwidth]{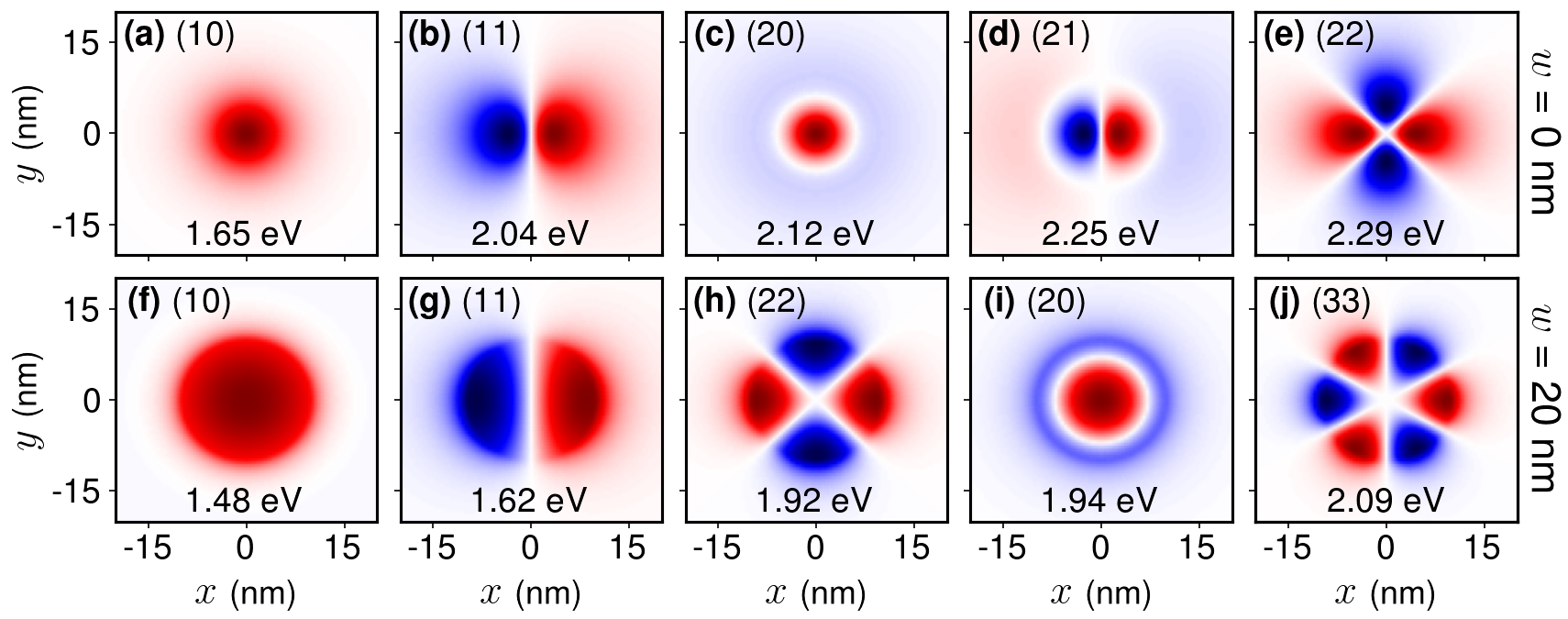}
\caption{The normalized QNM electric fields $E_{z,\ell\mathit{m}}$ in the horizontal xy-plane at the nanogap center for the first five QNMs of NPoMs with (a-e) $w=$ 0 nm and (f-j) $w=$ 20 nm. The real eigenfrequencies $\omega_{\ell\mathit{m}}$ in eV are shown at the bottom of all panels.}
\label{fig:qnmEz}
\end{figure}

Fig. \ref{fig:qnmEz}(a-e) further explores the field profiles of the first five QNMs in the spherical NPoM nanogap where the label ($\ell\mathit{m}$) is assigned to each QNM according to its symmetries. The label $\ell$ determines the number of nodes and antinodes along the radial coordinates $r=\sqrt{x^2 + y^2}$ whereas the label $\mathit{m}$ directly corresponds to the number of antinode pairs along the angular coordinate $\phi=\arctan(y/x)$ with $\widetilde{\mathbf{E}}_{\ell\mathit{m}}(r,\phi,z) \propto \exp(i\mathit{m}\phi)$. The ($\ell$0) modes have a circular symmetry. They also display antinodes at the center $r=0$ and $\ell - 1$ nodes along $r$. On the other hand, the QNMs with $\mathit{m} \neq 0$ have nodes at the center which extend radially at angles $(2n+1)\pi/2\mathit{m}$ for $n \in \mathbb{Z}$. For a given $\mathit{m} \neq 0$, each QNM with $\ell=\mathit{m}$ has only one node at the center, and each successive increment in $\ell$ gives one more node along $r$, as shown in Fig. \ref{fig:qnmEz}(b, d, e). Here, the QNMs with negative $\mathit{m} = -|\mathit{m}|$ are omitted since they are degenerate with those of the same $\ell$ but with positive $\mathit{m} = +|\mathit{m}|$. The field profile of the degenerate pair of each QNM can be easily obtained by rotating its fields by angle $\pi/2\mathit{m}$ around the z-axis.

Following this nomenclature, our calculations show that the first 20 QNMs of the spherical NPoM ($w=0$ nm), arranged in order of ascending real eigenfrequencies, are ($lm$) $=$ (10), (11), (20), (21), (22), (30), (31), (32), (33), (40), (41), (42), (43), (50), (44), (51), (52), (53), (60), (54), respectively. The results demonstrate two key features: (i) modes of the same $\ell$ have higher eigenfrequencies with increasing $\mathit{m}$, (ii) the first 13 QNMs with different $\ell$ do not intermix (see Fig. S2 in Supporting Information).

\section*{Faceted nanocavities}

Real metal nanoparticles are always faceted and lie on their facets with varying widths on a substrate \cite{sigle2015monitoring, benz2016sers}. Since the gap plasmonic resonances are highly sensitive to the gap morphology, the gold nanoparticles are modeled as truncated spheres with facet widths $w$ from 0 to 40 nm. The remaining nanoparticle facets outside the gap region do not play any significant role. We confirm that the nomenclature defined above is still preserved for faceted NPoMs although their resonant frequencies and optical properties are modified.

As the facet width increases, the optical environment at the nanogap approaches that of a plasmonic waveguide at large facet width. As a result, the order of QNMs with different $\ell$ start intermixing because the QNMs which are more waveguide-like, i.e. those with larger $\mathit{m}$, become more energetically favourable and tend to lie at lower eigenfrequencies \cite{tserkezis2015hybridization}. For example, the first 20 QNMs of a faceted NPoM with $w=20$ nm are ($lm$) $=$ (10), (11), (22), (20), (33), (21), (30), (44), (32), (31), (55), (40), (43), (66), (41), (42), (54), (77), (50), (51), respectively (Fig. S2 in Supporting Information illustrates this mode intermixing). The full list of the QNMs of NPoMs with $w =$ 0 - 40 nm is shown in Table S1 in Supporting Information. As the facet width increases from 0 to 20 nm, the field confinement becomes spread across the facet as shown in Figs. \ref{fig:qnm_E_xz}(d, e) and \ref{fig:qnmEz}(f-j) for the $w = $ 20 nm NPoM. Nonetheless, the symmetries of these QNMs are preserved for all facet widths, validating the nomenclature used in this article. 

Previous studies where the far-field spectra of NPoMs were analysed \cite{sigle2015monitoring, tserkezis2015hybridization, huh2018comparative, devaraj2019modifying} have identified only the ($\ell$0) and ($\ell$1) modes. The ($\ell$0) modes are bright and can be efficiently excited by the vertical (z-axis) component of an exciting plane wave. The ($\ell$1) modes, though much darker than the ($\ell$0), can be efficiently excited on NPoMs with either a horizontally polarized wave plane or two vertically polarized out-of-phase plane waves propagating along opposite directions \cite{demetriadou2017spatiotemporal}. On the other hand, higher-order QNMs with $|\mathit{m}| > 1$, such as (22) and (33), have not been reported in the literature before. These higher-order QNMs can not be easily isolated since they are very weakly coupled to external fields and spectrally overlap with each other. However, quantum emitters in the plasmonic nanocavities do couple to these QNMs, changing the far-field spectra observed. Hence, they are hidden in far-field spectra and have so far been mostly neglected.

In order to understand the optical behaviours of gap plasmonic resonances created by faceted NPoM cavities, it is useful to consider the high-order gap plasmonic resonances in the infinite frequency limit $\omega_{\ell\mathit{m}} \rightarrow \infty$, i.e. $\lambda_{\ell\mathit{m}} \rightarrow 0$. In this limit, the resonances no longer see the nanoparticle, and the system can simply be treated as a metal-insulator-metal (MIM) plasmonic waveguide \cite{tserkezis2015hybridization, mertens2016tracking}. The complex eigenfrequencies of the waveguide $\widetilde{\omega}_\mathrm{MIM} = \omega_\mathrm{MIM} - i\kappa_\mathrm{MIM}$ can be calculated by solving the semi-analytical parametric equation  \cite{maier2007plasmonics,baumberg2019extreme}
\begin{equation}
    \tanh \left( 
        d \sqrt{
            \beta^2 - (\widetilde{\omega}_\mathrm{MIM}/c)^2 ~ \epsilon_\mathrm{gap}
        } 
    \right) 
    = - \frac{
        \epsilon_\mathrm{gap}\sqrt{
            \beta^2 - (\widetilde{\omega}_\mathrm{MIM}/c)^2 ~ \epsilon_\mathrm{Au}(\widetilde{\omega}_\mathrm{MIM})
        }
    }{
        \epsilon_\mathrm{Au}(\widetilde{\omega}_\mathrm{MIM})\sqrt{
            \beta^2 - (\widetilde{\omega}_\mathrm{MIM}/c)^2 ~ \epsilon_\mathrm{gap}
        }
    }
    \label{eq:MIM}
\end{equation}
subject to the wavevector parameter $\beta$ whereas $d = $ 1 nm is the gap thickness,  $\epsilon_\mathrm{gap}=n^2_\mathrm{gap}$ is the gap permittivity and $\epsilon_\mathrm{Au}(\widetilde{\omega})$ is the gold permittivity. We note that the MIM plasmonic waveguide also has another set of solutions with the opposite parity to Eq. \ref{eq:MIM}. However, these solutions have zero $E_z$ field components at the gap center and do not correspond gap plasmons.

\begin{figure}[!ht]
\centering
\includegraphics[width=0.9\columnwidth]{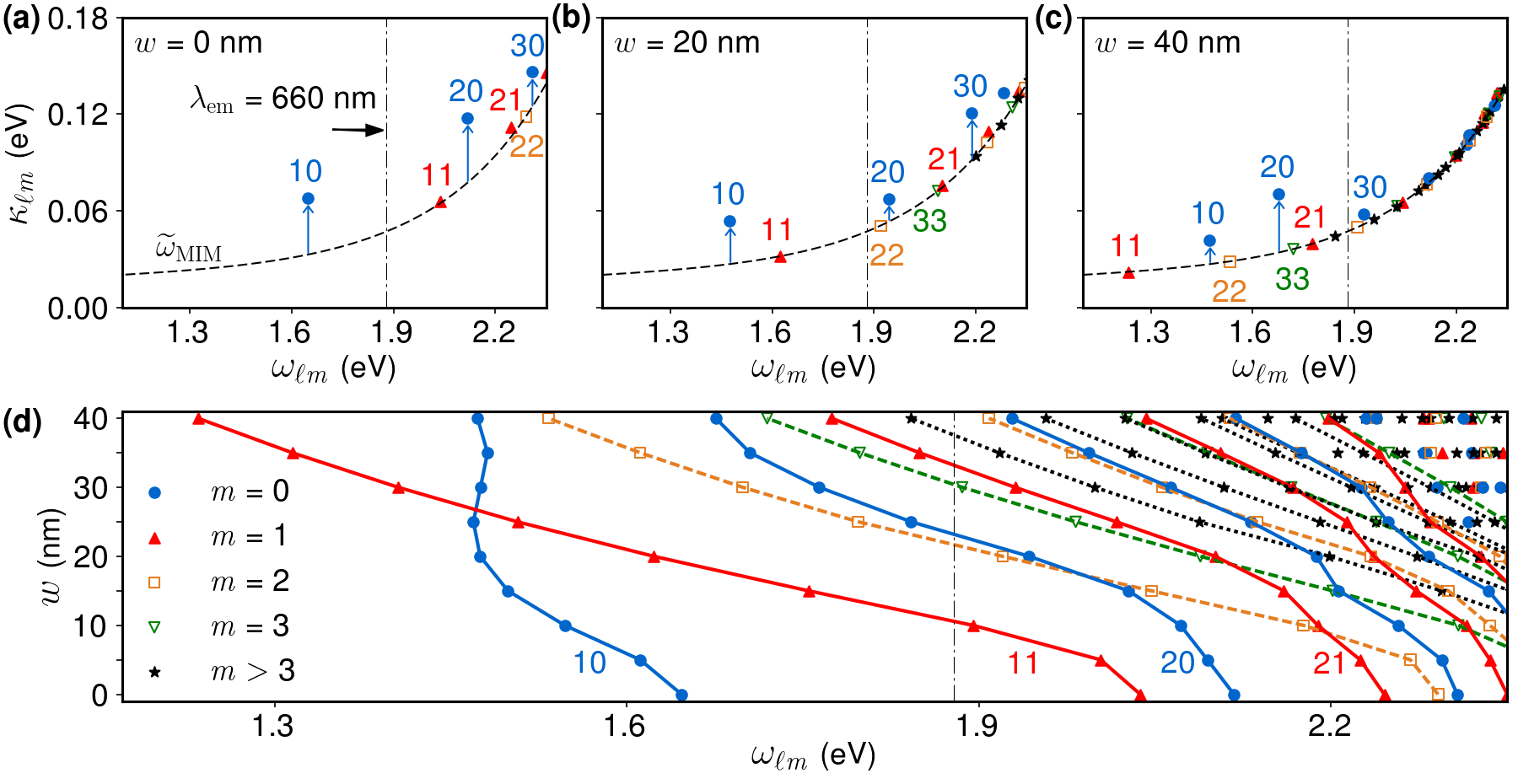}
\caption{The complex eigenfrequencies $\widetilde{\omega}_{\ell\mathit{m}} = \omega_{\ell\mathit{m}} - i\kappa_{\ell\mathit{m}}$ of NPoMs with facet widths (a) $w=0$ nm, (b) $w=20$ nm and (c) $w=40$ nm. The QNMs with $\mathit{m}=$ 0, 1, 2 and 3 are shown as blue circles, red triangles, orange squares and green inverted triangles, respectively, whereas $\mathit{m}>3$ QNMs are shown as black stars. The dashed lines correspond to the complex eigenfrequencies $\tilde{\omega}_\mathrm{MIM}$ of the MIM plasmonic waveguide whereas the dash-dotted vertical lines indicate the emitter wavelength $\lambda_\mathrm{em} =$ 660 nm (1.88 eV) which is considered in Figs. \ref{fig:alpha} and \ref{fig:nfft_x0}. The blue arrows indicate the radiative ($\ell$0) modes. (d) The real eigenfrequencies $\omega_{i}$ for $w=0-40$ nm. The fitted lines connect the QNMs with the same labels.} 
\label{fig:qnmFreq}
\end{figure}

Fig. \ref{fig:qnmFreq}(a-c) investigates the spectral correlations between the QNMs on complex eigenfrequency planes. We plot with a dashed line the solution of Eq. \ref{eq:MIM}, which corresponds to the waveguide eigenfrequencies $\widetilde{\omega}_\mathrm{MIM}$, as well as the resonant frequencies $\widetilde{\omega}_{\ell\mathit{m}}$ of the NPoMs. For all facet widths, all QNMs with $\mathit{m}>0$ lie close to the dashed line of $\widetilde{\omega}_{\ell\mathit{m}}$. As $w$ increases, the QNMs with $\mathit{m}>0$ simply migrate along $\widetilde{\omega}_\mathrm{MIM}$ to lower real-eigenfrequencies. Since the imaginary eigenfrequencies $\kappa_{\ell\mathit{m}}$ represent the energy dissipation of the QNMs, these results demonstrate that those with $\widetilde{\omega}_{\ell\mathit{m}}$ lying along $\widetilde{\omega}_\mathrm{MIM}$ are dark modes, and their energies are dissipated almost exclusively through heat at the rate $\kappa_\mathrm{MIM}$. Only the ($\ell$0) modes are efficiently radiative and lie significantly above $\widetilde{\omega}_\mathrm{MIM}$ because their charge distributions have non-zero electric dipole moments and are therefore unlike the modes predicted by the MIM model. This implies that the differences between $\kappa_{\ell\mathit{0}}$ and $\kappa_\mathrm{MIM}$ correspond to energy dissipation through far-field radiation (shown as blue arrows in Fig. \ref{fig:qnmFreq}(a-c)).

To visualize how the QNMs evolve with varying facet width $w$, Fig. \ref{fig:qnmFreq}(d) plots real eigenfrequencies $\omega_{\ell\mathit{m}}$ of all QNMs of NPoMs with facet width $w$ between 0 and 40 nm. Most QNMs red-shift with increasing $w$. There are a few exceptions such as the (10) modes which start at $w=$ 0 nm as the lowest eigenfrequency QNM and then level off beyond $w>$ 25 nm. The QNMs with larger $\mathit{m}$ generally red-shift at more rapid rates. For example, the (11) mode appears at lower frequency than the (10) mode for facet sizes around $w=$ 25 nm, and the (22) mode becomes more energetically favorable than the (21) and (20) modes for facet sizes $w=$ 10 and 15 nm, respectively. These results demonstrate complicated spectral relations between bright and dark resonances of NPoMs with different facet widths, leading to vastly different near- and far-field optical behaviors.

\begin{figure}[!ht]
\centering
\includegraphics[width=0.6\columnwidth]{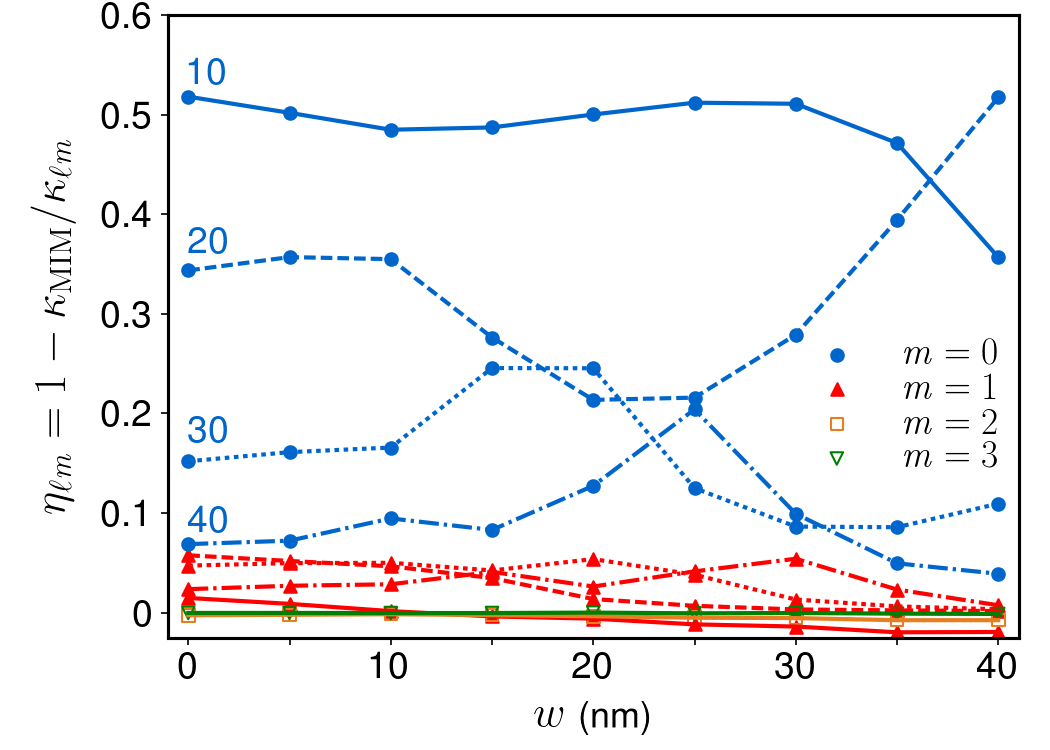}
\caption{Radiation efficiencies $\eta_{\ell\mathit{m}}$ of (10), (20), (30), (40), (11), (21), (31), (41), (22) and (33) modes of NPoMs with facet width from 0 to 40 nm. The ($\ell$0), ($\ell$1), (22) and (33) modes are shown as blue, red, orange and green lines, respectively. The ($\ell$0) and ($\ell$1) modes are also further distinguished by solid lines for $\ell=1$, dashed lines for $\ell=2$, dotted lines for $\ell=3$ and dashed-dotted lines for $\ell=4$.}
\label{fig:rad_eff}
\end{figure}

To further explore the radiative nature of each QNM, we estimate the radiation efficiency of the ($\ell\mathit{m}$) mode as
\begin{equation}
    \eta_{\ell\mathit{m}} = \frac{\kappa_{\ell\mathit{m}}-\kappa_\mathrm{MIM}}{\kappa_{\ell\mathit{m}}} = 1 - \frac{\kappa_\mathrm{MIM}}{\kappa_{\ell\mathit{m}}}.
\end{equation}
Figure \ref{fig:rad_eff} shows the radiation efficiencies of (10), (20), (30), (40), (11), (21), (31), (41), (22) and (33) modes. Overall, the ($\ell$0) modes are the dominant radiative channels of NPoMs, as one would expect. The (10) mode has the highest efficiency for a wide range $w$ and is only overtaken by the (20) mode for $w > 35$ nm. As $\ell$ increases, the ($\ell$0) modes becomes less radiative at $w =$ 0 nm but show complex behaviors near $w =$ 20 nm. The ($\ell$1) modes also have non-negligible but small radiation efficiencies below 0.1. In fact, these ($\ell$1) modes take an essential part in determining the far-field emission, as later shown in Figs. \ref{fig:alpha} and \ref{fig:nfft_x0}. On the other hand, the modes with $\mathit{m}>1$ are virtually dark, and their radiation efficiencies are close to zero. We note that the efficiencies of some QNMs, such as the (11) and (22) modes, fall below zero at large facet width. This is because the MIM plasmonic waveguide represents the NPoM system phenomenologically and, hence, only provides an approximate description.

\section*{Far-field emission profiles}

\begin{figure}[!ht]
\centering
\includegraphics[width=1.0\columnwidth]{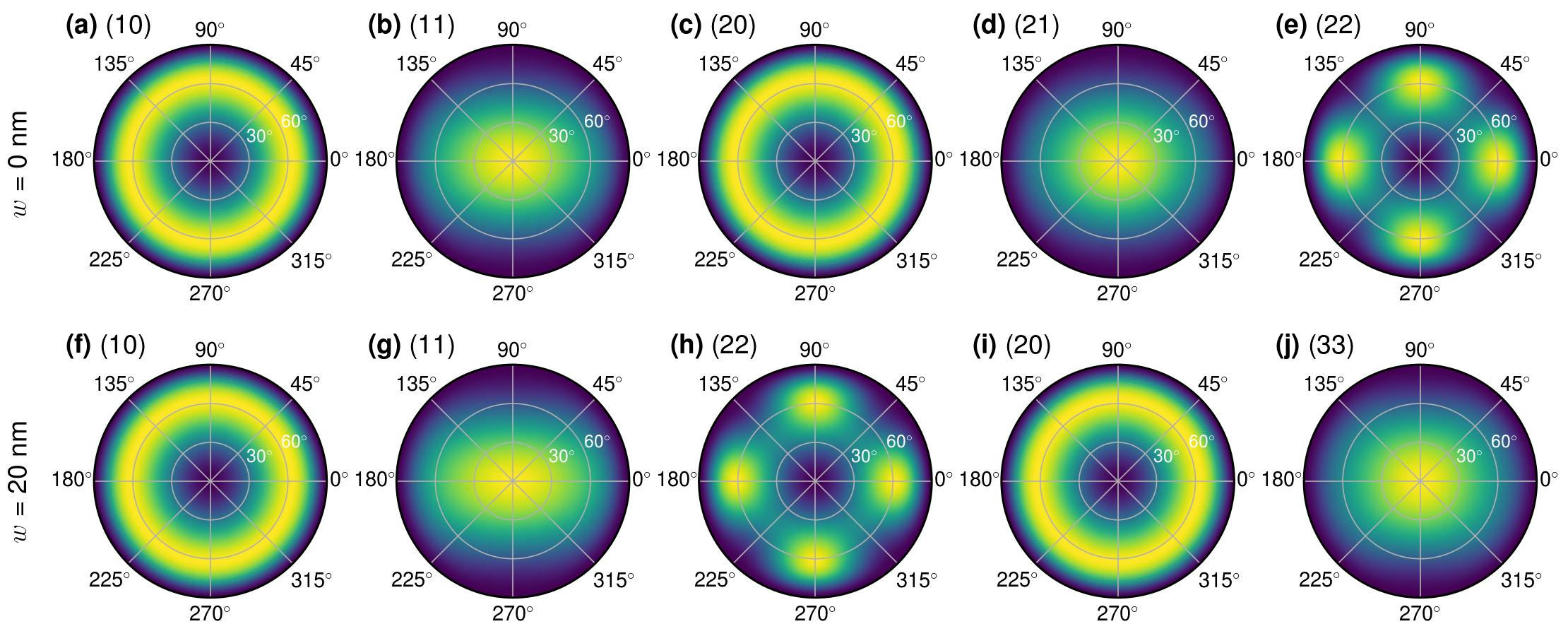}
\caption{The normalized Poynting fluxes $\langle S_{\ell\mathit{m}} \rangle$ for the first five QNMs on the top hemisphere above NPoMs with (a-e) $w=$ 0 nm and (f-j) $w=$ 20 nm.}
\label{fig:nfft_qnm}
\end{figure}

Having investigated the near-field profiles and eigenfrequencies of individual eigenmodes for NPoMs, it is now straightforward to examine their respective far-field properties. We employ the software code RETOP \cite{yang2016near} which implements a NFFT transformation in dispersive stratified media \cite{balanis2016antenna, demarest1996fdtd, pors2015quantum}. We project the electromagnetic fields of each QNM in the near-field zone, $(\widetilde{\mathbf{E}}_{\ell\mathit{m}},\widetilde{\mathbf{H}}_{\ell\mathit{m}})$, to the far-field, $(\widetilde{\mathbf{E}}_{\ell\mathit{m}}^{\mathrm{ff}},\widetilde{\mathbf{H}}_{\ell\mathit{m}}^{\mathrm{ff}})\exp(ik_{i} R_u)$ where $k_{i} = \omega_{\ell\mathit{m}}/c$, on the upper hemisphere of radius $R_u$ above the NPoMs. The time-average Poynting flux of each QNM, $\langle S_{\ell\mathit{m}} \rangle = \mathrm{Re}[(\widetilde{\mathbf{E}}^{\mathrm{ff}}_{\ell\mathit{m}})^{*}\times\widetilde{\mathbf{H}}^{\mathrm{ff}}_{\ell\mathit{m}}]/2$, is then evaluated and shown in Fig. \ref{fig:nfft_qnm} for spherical ($w =$ 0 nm) and faceted ($w =$ 20 nm) NPoMs. The bright ($\ell$0) modes show ring-shaped emission patterns with emission peaks near angle $\theta=$ 60$^\circ$ while the darker ($\ell$1) modes show spot-shaped emission, peaked at $\theta=$ 0$^\circ$. These results are consistent with those reported previously in Ref. \cite{mubeen2012plasmonic, chikkaraddy2017ultranarrow, baumberg2019extreme}. The ($22$) modes in Fig. \ref{fig:nfft_qnm}(e, h) shows far-field emission with four emission lobes, having the same symmetry as their near-field profiles in Fig. \ref{fig:qnmEz}(e, h). On the other hand, the ($33$) mode in Fig. \ref{fig:nfft_qnm}(j) shows spot-shaped emission, similar to those of the ($\ell$1) modes.

Although it is indeed essential to understand how each QNM radiates to the far-field, individual QNMs are rarely excited in isolation. For example, an emitter placed in a NPoM nanocavity couples to a specific collection of the modes, depending on its transition frequency. Hence, accurate far-field analysis of a NPoM must involve reconstructing the collective QNM far-field emission. Here, the coupling coefficients of an electric dipole emitter placed inside a NPoM to the QNMs are evaluated and used to calculate the total far-field emission of the dipole emitter.

For a single emitter with dipole moment $\boldsymbol{\mu}$ and transition frequency $\omega_\mathrm{em}$ placed at position $\mathbf{r}_\mathrm{em}$ in close proximity to a plasmonic structure, the electromagnetic fields $(\mathbf{E},\mathbf{H})$ radiated by the emitter can be expanded to a small set of QNMs 
\begin{equation}
    \mathbf{E}(\mathbf{r}) = \sum_{i} \alpha_{i}(\mathbf{r}_\mathrm{em};\omega_\mathrm{em})\widetilde{\mathbf{E}}_{i}(\mathbf{r})
\label{eq:QNMdecomposition}
\end{equation}
where $\alpha_{i}$ is the modal excitation coefficient which can be calculated by \cite{sauvan2013theory}
\begin{equation}
    \alpha_{i}(\mathbf{r}_\mathrm{em};\omega_\mathrm{em}) = - \omega\sum_{j}B_{i,j}^{-1} (\omega_\mathrm{em})\boldsymbol{\mu}\cdot\widetilde{\mathbf{E}}_{j}(\mathbf{r}_\mathrm{em}).
\label{eq:qnmAlpha}
\end{equation}
\begin{equation}
    B_{i,j}(\omega) = 
    \iiint_\Omega \left\{
        \widetilde{\mathbf{E}}_{j}\cdot[\omega\varepsilon(\mathbf{r};\omega) - \widetilde{\omega}_{i}\varepsilon(\mathbf{r};\widetilde{\omega}_{i})]\widetilde{\mathbf{E}}_{i}
        - \mu_0\widetilde{\mathbf{H}}_{j}\cdot(\omega-\widetilde{\omega}_{i})\widetilde{\mathbf{H}}_{i}        
    \right\} d\mathbf{r}^3.
\label{eq:qnmB}
\end{equation}

\begin{figure}[!ht]
\centering
\includegraphics[width=0.7\columnwidth]{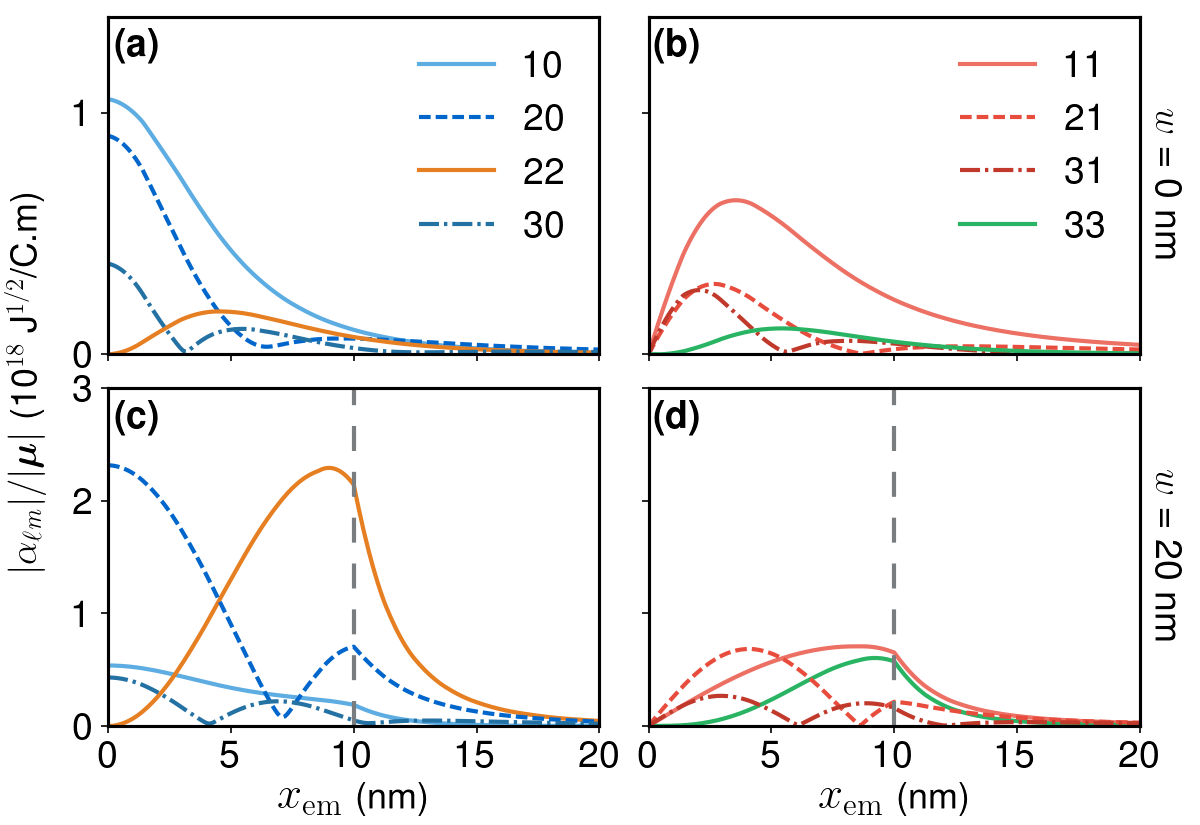}
\caption{QNM coupling coefficient $\alpha_{\ell\mathit{m}}$ to an emitter with dipole moment $\boldsymbol{\mu}$ and transition wavelength $\lambda_\mathrm{em} = 660$ nm for different positioning $x_\mathrm{em}$ of the emitter. (a, c) First four symmetric and (b, d) first four antisymmetric QNMs of NPoMs with facet widths (a, b) $w=0$ nm and (c, d) $w=20$ nm. The vertical dashed lines indicate the facet edges.}
\label{fig:alpha}
\end{figure}

Here, a methylene blue molecule with transition wavelength of $\lambda_\mathrm{em}=$ 660 nm is chosen as a quantum emitter of choice since it is a typical dye molecule experimentally used inside a NPoM \cite{chikkaraddy2016single}. By placing the emitter with $\boldsymbol{\mu} = |\boldsymbol{\mu}|\hat{\mathbf{e}}_z$ at position $x_\mathrm{em}$ in the gap of a NPoM, its coupling to the NPoM's QNMs in Eq. (\ref{eq:qnmAlpha}) can be simplified to
\begin{equation}
    \alpha_{\ell\mathit{m}}(x_\mathrm{em};\omega_\mathrm{em})/|\boldsymbol{\mu}| = - \omega_\mathrm{em}\sum_{\ell'\mathit{m}'}B_{\ell\mathit{m},\ell'\mathit{m}'}^{-1} (\omega_\mathrm{em})\widetilde{E}_{z,\ell'\mathit{m}'}(x_\mathrm{em})
\end{equation}
where $\widetilde{E}_{z,i} = \hat{\mathbf{e}}_z\cdot\widetilde{\mathbf{E}}_{i}$, $\omega_\mathrm{em}=2\pi c/\lambda_\mathrm{em}$, and only the first 20 QNMs are included in the calculations, as shown in Table S1. The coupling coefficients $|\alpha_{\ell\mathit{m}}|$ to different ($\ell\mathit{m}$) modes are shown in Fig. \ref{fig:alpha}, as the emitter moves laterally inside the gap along the x-axis. Four QNMs with even $\mathit{m}$, Fig. \ref{fig:alpha}(a, c), are separated from those with odd $\mathit{m}$, Fig. \ref{fig:alpha}(b, d), as they have symmetric and antisymmetric field profiles across the $x=$ 0 nm plane, respectively. For $w=$ 0 nm, the emitter's resonance lies spectrally close to the (10), (11) and (20), see Fig. \ref{fig:qnmFreq}(a). In Figs. \ref{fig:alpha}(a, b), the coupling magnitudes of these three QNMs indeed dominate the coupling with the emitter. The coupling magnitudes of different QNMs change drastically when the facet width increases to $w=$ 20 nm. The emitter instead lies spectrally close to (11), (22), (20), (33) and (21) see  Fig. \ref{fig:qnmFreq}(b), which become the modes coupled to the emitter, as shown in Fig. \ref{fig:alpha}(c, d). The relative magnitudes of $|\alpha_{\ell\mathit{m}}|$ are crucial in understanding the total far-field emission.

\begin{figure}[!ht]
\centering
\includegraphics[width=0.95\columnwidth]{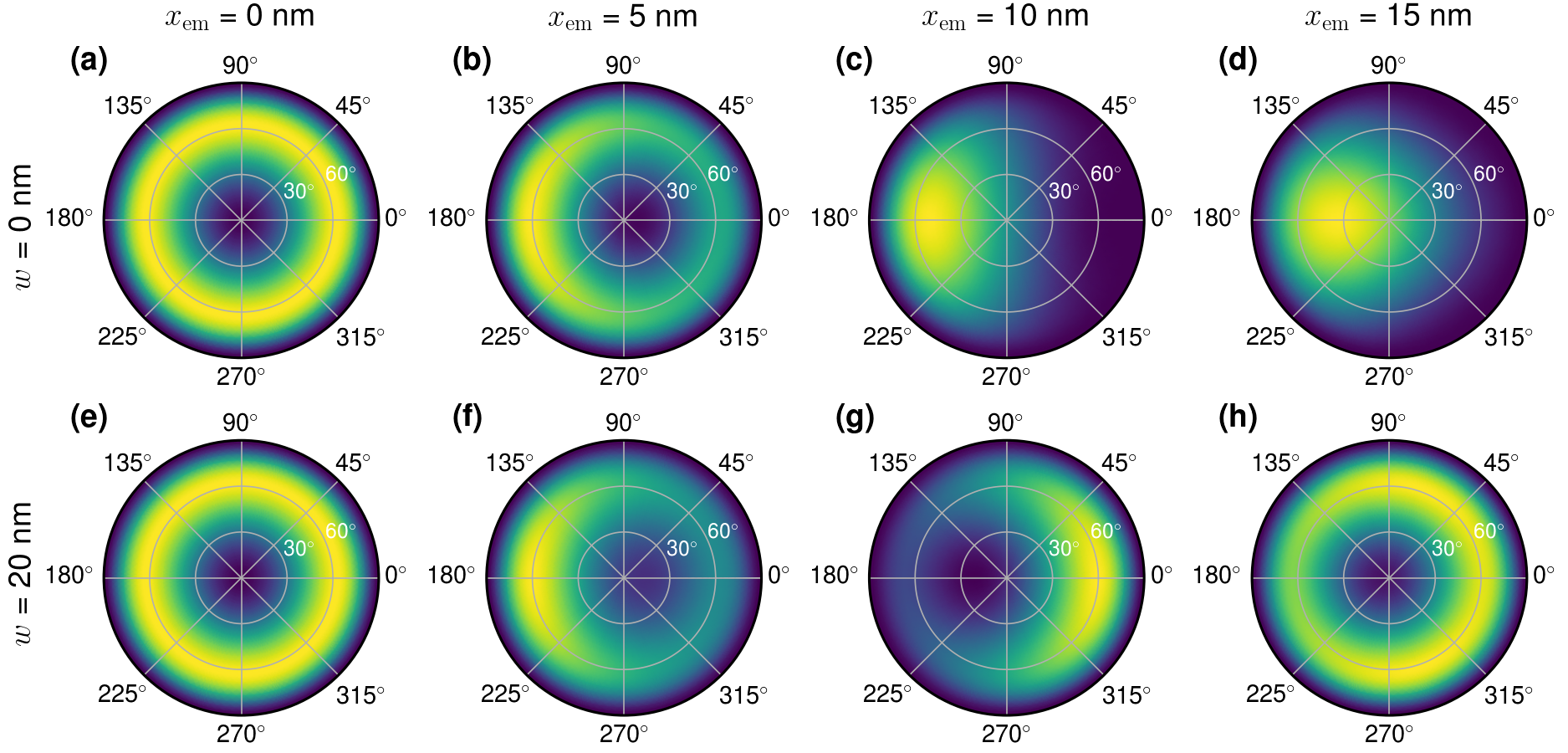}
\caption{The normalized total Poynting fluxes $\langle S_\mathrm{tot} \rangle$ of dipole emission on the top hemispheres above NPoMs, obtained by the QNM method. Single emitters with transition wavelength $\lambda_\mathrm{em} = 660$ nm are placed in the nanogap with facet width (a-d) 0 nm and (e-h) 20 nm at position (a, e) $x_\mathrm{em} = 0$ nm, (b, f) $x_\mathrm{em} = 5$ nm, (c, g) $x_\mathrm{em} = 10$ nm and (d, h) $x_\mathrm{em} = 15$ nm.}
\label{fig:nfft_x0}
\end{figure}

The total far-field emission $(\widetilde{\mathbf{E}}^{\mathrm{ff}}, \widetilde{\mathbf{H}}^{\mathrm{ff}})$ from the emitter placed inside a NPoM can then be evaluated by combining the far-field QNM fields $(\widetilde{\mathbf{E}}_{\ell\mathit{m}}^{\mathrm{ff}},\widetilde{\mathbf{H}}_{\ell\mathit{m}}^{\mathrm{ff}})$ with complex coupling coefficient $\alpha_{\ell\mathit{m}}$
\begin{equation}
    (\widetilde{\mathbf{E}}^{\mathrm{ff}}(x_\mathrm{em};\omega_\mathrm{em}),\widetilde{\mathbf{H}}^{\mathrm{ff}}(x_\mathrm{em};\omega_\mathrm{em})) = \sum_{\ell\mathit{m}}\alpha_{\ell\mathit{m}}(x_\mathrm{em};\omega_\mathrm{em})(\widetilde{\mathbf{E}}_{\ell\mathit{m}}^{\mathrm{ff}},\widetilde{\mathbf{H}}_{\ell\mathit{m}}^{\mathrm{ff}}).
\end{equation}
Fig. \ref{fig:nfft_x0} shows the total time-average Poynting flux $\langle S_\mathrm{tot} \rangle = \mathrm{Re}[(\widetilde{\mathbf{E}}^{\mathrm{ff}})^{*}\times\widetilde{\mathbf{H}}^{\mathrm{ff}}]/2$ from the emitter placed at lateral position $x_\mathrm{em}=$ 0, 5, 10 and 15 nm inside NPoMs with facet width $w=$ 0 and 20 nm. For $w=$ 0 nm, Fig. \ref{fig:nfft_x0}(a-d), the emission has a ring-shaped pattern at $x_\mathrm{em}=$ 0 nm as the emitter is dominantly coupled to the (10) and (20) modes. As the emitter is moved away from the nanocavity centre, the emitter couples more efficiently to the (11) mode. The spot-shaped emission from the (11) mode also has a phase variation over the angular coordinate $\phi$. At $x_\mathrm{em}=$ 5 nm, the distorted ring-shaped emission in Fig. \ref{fig:nfft_x0}(b) is the result of this phase variation, giving destructive interference for $-\pi/2 < \phi < \pi/2$ and constructive interference for $\pi/2 < \phi < 3\pi/2$ with the ring-shaped emission from the (10) and (20). For larger $x_\mathrm{em}\geq$ 10 nm, the antisymmetric (11) mode dominates the coupling with the emitter, and the emission becomes an offset spot-shaped pattern.

For the facet NPoM with facet $w=$ 20 nm, the emission also has a ring-shaped pattern when the emitter is at the centre of the cavity since the emitter couples mainly to the (20) mode, see Fig. \ref{fig:nfft_x0}(e). By moving the emitter within the faceted NPoM to position $x_\mathrm{em}=$ 5 nm the far-field emission becomes a distorted ring-shaped pattern since the emitter couples more efficiently to the (11) and (21) modes, which is similar to the spherical NPoM in Fig. \ref{fig:nfft_x0}(b). However, when the emitter moves even closer to the facet edge at $x_\mathrm{em}=$ 10 nm the far-field emission pattern flips 180$^\circ$, as shown in Fig. \ref{fig:nfft_x0}(g). 
This flip results from changes in the complex coefficients $\alpha_{\ell\mathit{m}}$ of different QNMs which instead interferes constructively for $-\pi/2 < \phi < \pi/2$ and destructively for $\pi/2 < \phi < 3\pi/2$. Unlike Fig. \ref{fig:nfft_x0}(d), the emission at $x_\mathrm{em}>$ 15 in Fig. \ref{fig:nfft_x0}(h) resumes the ring-shaped pattern for the emitter since the emitter dominantly couples the symmetric (20) and (22) modes. 

The far-field emission profiles in Fig. \ref{fig:nfft_x0} are quantitatively calculated by combining 20 QNMs. In fact, only a small number of dominant QNMs are required in order to qualitatively capture the main contributions to the far-field profiles. As shown in Fig. S3 in Supporting Information, the (10) and (11) modes are the only dominant QNMs for the 660 nm emitter in the $w = 0$ nm NPoM. On the other hand, the emission from the faceted NPoM with $w =$ 20 nm requires up to eight QNMs, including the (10), (20), (30), (11), (21), (31), (41) and (51) modes. These results are also confirmed by independent calculations in Fig. S4 in Supporting Information where the far-field emission is calculated directly from an electric dipole source placed in the NPoMs, effectively exciting all available QNMs. Excellent agreements between Fig. \ref{fig:nfft_x0}, S3 and S4 conclude that our calculations involving the first 20 QNMs provide a sufficient quantitative description of the NPoMs.

Fig. \ref{fig:nfft_x0} demonstrates multi-modal interaction between the NPoM resonances and the emitter which lead to counter-intuitive far-field emission patterns, depending on the NPoM gap morphology and the emitter's lateral position. These results are essential for understanding experimental measurements involving the emission from emitters placed inside a plasmonic nanocavity and deciphering to which plasmon modes they actually interact with.

\section*{Conclusions}
In recent years, the NPoM geometry has become a prominent nanostructure in nanoplasmonics due to its extreme light confinement properties. However, a comprehensive modal analysis of the structure had not in general been available, and most current studies infer its resonances from far-field optical scattering. We use the quasinormal mode (QNM) approach to analyze the morphology-dependent plasmonic resonances of a NPoM structure. A collection of bright and dark resonances are revealed, some of which have not yet been previously reported. A simple and comprehensive nomenclature is introduced based on spherical harmonics which reflects the underlying charge distributions on the nanoparticles. The near-field and far-field optical behaviors of NPoMs with varying facet widths are also reported, which clarifies the inconsistency in previous near-field and far-field analyses. This study also unveils rich and intricate multi-modal interactions with a single quantum emitter and has the potential to aid the design of quantum plasmonic experiments, such as quantum computing with DNA-origami-controlled qubits \cite{chikkaraddy2018mapping} and quantum plasmonic immunoassay sensing \cite{kongsuwan2019quantum}.

\begin{acknowledgement}

We acknowledge support from the Engineering and Physical Sciences Research Council (EPSRC) UK through projects EP/L024926/1, EP/L027151/1, EP/N016920/1 and NanoDTC EP/L015978/1. AD acknowledges support from the Royal Society URF/R1/180097 and RGF/EA/181038. RC acknowledges support from Trinity College Cambridge.

\end{acknowledgement}





\providecommand{\latin}[1]{#1}
\makeatletter
\providecommand{\doi}
  {\begingroup\let\do\@makeother\dospecials
  \catcode`\{=1 \catcode`\}=2 \doi@aux}
\providecommand{\doi@aux}[1]{\endgroup\texttt{#1}}
\makeatother
\providecommand*\mcitethebibliography{\thebibliography}
\csname @ifundefined\endcsname{endmcitethebibliography}
  {\let\endmcitethebibliography\endthebibliography}{}

\end{document}